\documentclass{class}
\usepackage{amsmath,bm,mathrsfs,xcolor}
\usepackage{graphicx}
\usepackage{epstopdf, epsfig,psfrag}
\usepackage{cases,pdfpages}
\usepackage[normalem]{ulem}

\shorttitle{Effect of polymer-stress diffusion in the
numerical simulation of elastic turbulence}
\shortauthor{A. Gupta and D. Vincenzi}

\title{Effect of polymer-stress diffusion in the
numerical simulation of elastic turbulence}

\author{
Anupam Gupta\aff{1}
 \and 
Dario Vincenzi\aff{2}
  \corresp{\email{dario.vincenzi@unice.fr}}
}

\affiliation{
\aff{1}Paulson School of Engineering and Applied Sciences, Harvard University, Cambridge, Massachusetts 02138, USA
\aff{2}Universit\'e C\^ote d'Azur, CNRS, LJAD, 06100 Nice, France
}

\begin{document}

\maketitle

\begin{abstract}
Elastic turbulence is a chaotic regime that emerges 
in polymer solutions at low Reynolds numbers.
A common way to ensure stability in numerical simulations
of polymer solutions
is to add artificially large polymer-stress diffusion.
In order to assess the accuracy of this approach in the elastic-turbulence
regime,
we compare numerical simulations of the two-dimensional
Oldroyd-B and FENE-P models sustained by 
a cellular force with and without artificial diffusion.
We find that
artificial diffusion can have a dramatic effect even
on the large-scale properties of the flow and we show
some of the spurious phenomena that may arise when artificial
diffusion is used.
\end{abstract}


\section{Introduction}

Polymer solutions can exhibit a chaotic behaviour, even at low
Reynolds numbers, as a result of purely elastic instabilities \citep{GS00,GS04}.
This regime of polymer solutions, known as elastic turbulence,
finds a natural application in
microfluidics, where the typical Reynolds numbers are very low and the
addition of polymers to the fluid can be
employed to accelerate phenomena such as mixing \citep{GS01}, 
emulsification \citep{poole12} and heat transfer \citep*{TCB15,poole16}
or to examine the dynamics of microscopic objects in 
fluctuating flows \citep{LS14}.

The numerical simulation of elastic turbulence is challenging for at least three reasons.
The first two are common to the high-Reynolds-number regime of polymer
solutions.
Indeed,
the available constitutive
models of viscoelastic fluids are based on rather crude approximations and
disregard several, potentially relevant aspects of polymer dynamics
\citep{BHAC87}.
In addition, a high spatial resolution and advanced numerical schemes are
needed to resolve the sharp gradients that form in the polymer-stress field \citep{JC07,PGVG17}.
The third reason is specific to elastic turbulence and is the requirement that the time step used for the integration of the Navier--Stokes equations 
be small; 
this requirement is a consequence of the 
high viscosity of the fluid \citep{numerical_recipes}.

Most numerical studies of elastic turbulence
have therefore considered simplified flow configurations 
(e.g. in two dimensions and/or with
periodic boundary conditions) and have been restricted to
limited ranges of parameters.
Notwithstanding, even simple models of viscoelastic fluids in idealized settings
have proved successful in capturing the main properties of elastic turbulence.
Several experimental observations
are reproduced, at least qualitatively, by the Oldroyd-B model,
in which only the slowest oscillation mode of the polymer is retained
and polymer elasticity is assumed to be linear
\citep*{BBBCM08,TS09,BB10,TST11,GVQE13,PGVG17,vBSS18,GCMB18}.
Other studies have used the FENE-P model, which improves on the Oldroyd-B one by
taking into account the finite extensibility of polymers
\citep*{LK13,GP17}, or have taken a
Lagrangian approach and have solved
the dynamics of a large number of dumbbell-like polymers \citep{WG13,WG14}.
In fact, even a low-dimensional `shell model' of viscoelastic 
fluid reproduces elastic turbulence qualitatively \citep{RV16}.

Constitutive models
such as the Oldroyd-B and the FENE-P ones consist of a coupled
system of partial differential equations for the velocity of the solution
and for the polymer stress tensor.
This latter is by nature positive definite,
but numerical errors may lead to the loss of this property
and hence to instabilities \citep{J90}.
A standard way to prevent such instabilities is
to include {\it global} artificial diffusivity
in the model, i.e. to add a Laplacian term to the
evolution equation for the polymer stress with a space-independent coefficient \citep{SB95}.
Numerical simulations of turbulent polymer solutions that use 
artificial diffusivity are in qualitative agreement with experiments
\citep[e.g.][and references therein]{G14}.
Polymer-stress diffusion in fact has a physical origin, for it results from the diffusion of the
centre of mass of polymers due to thermal noise \citep{EK89}. However,
the values of diffusivity needed to achieve numerical stability 
are three to six orders of magnitude greater than those appropriate for real polymers
\citep[e.g.][]{VRBC06}.
For this reason, numerical schemes have been proposed that
avoid using  artificial diffusivity. 
These include, inter alia, 
schemes that only employ polymer diffusivity at those locations in the fluid where the polymer stress
loses its positive definite character \citep*{MYC01},
methods adapted from hyperbolic solvers \citep{VRBC06},
or schemes based on 
representations of the polymer stress tensor that guarantee the preservation
of its positive definiteness \citep*{VC03,FK04,BTRD11,zaki18}.
Such numerical schemes have been compared with simulations
using  artificial diffusivity at high or moderate Reynolds numbers,
and quantitative 
discrepancies have emerged: for instance, the level of drag reduction is diminished
by  artificial diffusivity,
the velocity and polymer-stress fields are significantly smeared,
excessive polymer-stress diffusion
leads to relaminarization (\citealt{MYC01}; \citealt{VRBC06}; \citealt*{STD18}).
Thus a consensus seems to have formed that 
at high or moderate Reynolds numbers
alternative methods should be preferred to the use of 
artificial diffusivity. 

At low Reynolds numbers, several studies on elastic turbulence have employed
 artificial diffusivity~\citep{BBBCM08,TS09,BB10,TST11,LK13,GCMB18}.
It was shown in \cite{T11} that in viscoelastic creeping flows
this has the effect of smoothing the
polymer-stress field and keeping it bounded.
However, to the best of our knowledge, the effect of artificial diffusivity
in the elastic-turbulence regime has not been examined yet.
This is the object of the present study, in which
we compare numerical simulations
with and without artificial diffusivity. As a case study,
we consider the Oldroyd-B model with a cellular forcing on a periodic square
(analogous simulations of the FENE-P model are presented
in the Appendix).
The cellular
forcing generates distinct regions of straining and vorticity
and thus allows us to describe a flow configuration in which the effect of artificial diffusivity
is particularly adverse.
Our results demonstrate that
the properties of the velocity field are strongly affected,
to such an extent that also the large-scale flow may be misrepresented.
In particular, we show that some phenomena observed in previous simulations
are due to artificial stress diffusion and are not present when alternative
integration methods are used.

The general effect of artificial diffusivity is to spread high polymer stresses
over large regions of the flow, including those where the polymer stress
would be weak because vorticity dominates and polymers should not be stretched.
We shall se that, 
in elastic turbulence, this fact leads to a spurious symmetry breaking
analogous to that observed in \cite{TS09} and \cite{TST11}.
Indeed, in low-strain regions polymers are weakly stretched and the external
force
would naturally impose its spatial structure, but if high polymer
stresses artificially diffuse into those regions, they prevail over the force
and provoke a strong modification of the
large-scale flow. This dynamics is
specific to elastic turbulence, because the Reynolds number is low and, in the absence
of artificial diffusivity,
the original laminar flow 
is weakly perturbed by the addition of polymers in vorticity-dominated regions.
At high Reynolds numbers the effect of artificial diffusivity is less dramatic, 
since the flow is already
chaotic before the addition of polymers and therefore
artificial diffusivity does
not change the large-scale flow so strongly.
Indeed, at high Reynolds numbers,
the differences between simulations with and without artificial diffusivity
are essentially quantitative \citep{VRBC06,STD18}.
In conclusions, we show that in elastic turbulence
artificial diffusivity induces dramatic qualitative
modifications of the large-scale flow,
which are not observed at high Reynolds numbers. Hence the study of the
low-Reynolds-number regime requires a separate study.

\section{Viscoelastic model}
\label{sect:model}

The Oldroyd-B model \citep{O50}  describes the deformation of polymers
by means of a space-time dependent positive-definite 
tensor: the polymer
conformation tensor $\mathsfbi{C}(\bm x,t)$.
In the limit of vanishing inertia, the coupled
evolution of $\mathsfbi{C}(\bm x,t)$ and the velocity field $\bm u(\bm x,t)$
that describes the motion of the
solution is given by the following equations:
\begin{subequations}
\begin{equation}
\label{eq:stokes}
\bnabla p=\nu \Delta\bm u +\dfrac{\mu}{\tau}\,\bnabla\bcdot\mathsfbi{C}+\bm f,
\qquad \bnabla\bcdot\bm u=0,
\end{equation}
\begin{equation}
\label{eq:conformation}
\partial_t \mathsfbi{C}+\bm u\bcdot\bnabla\mathsfbi{C}=(\bnabla\bm u)\bcdot\mathsfbi{C}
+\mathsfbi{C}\bcdot\bnabla\bm u)^\top-\dfrac{1}{\tau}(\mathsfbi{C}-\mathsfbi{I}),
\end{equation}
\label{eq:OB}
\end{subequations}%
where $p$ is pressure, $\nu$ is the kinematic viscosity of the solvent,
$\tau$ is the polymer relaxation time,
the components of the velocity gradient are defined as
$(\bnabla\bm u)_{ij}=\partial u_i/\partial x_j$ and $\mathsfbi{I}$ is the identity matrix.
The coupling coefficient $\mu$
represents the polymer contribution to the total kinematic viscosity of the solution
and is proportional to the concentration of polymers.
The body force $\bm f(\bm x)$ that sustains the  
motion of the solution is such that $\bnabla\bcdot\bm f=0$.
In the above equations, the conformation tensor is rescaled with
the polymer mean square equilibrium extension in the absence of flow.

In considering the limit of the Oldroyd-B model for vanishing inertia,
we follow \cite{FL03,TS09,TST11,BTRD11}, who describe the motion of the
solution by means of the Stokes equations in lieu 
of the Navier--Stokes equations.
However, as is discussed in \S~\ref{sect:conclusions}, our conclusions 
on the effect of artificial diffusivity are
unchanged if we use the Navier--Stokes equations with a Reynolds number
smaller than the critical value for the appearance of inertial instabilities.

Equations~\eqref{eq:OB} are studied on the two-dimensional
domain $V=[0,2\pi]^2$ with periodic boundary
conditions. We consider the cellular force:
\begin{equation}
\bm f(\bm x)= f_0(-\sin Ky, \sin Kx),
\end{equation}
where $f_0$ is the amplitude and $K$ the spatial frequency.
Cellular-like forcings have been widely used in
experiments of chaotic mixing in two-dimensional flows \citep*[e.g.][]{CMT94,RHG99}.
As is discussed in the conclusions, the effect of artificial
diffusivity for other forcings may not be equally dramatic;
however, the cellular forcing allows us to clearly identify the 
fashion in which artificial diffusivity operates in elastic turbulence
and to demonstrate how strong its effect can be in this regime.

\begin{figure}
\centering
\includegraphics[width=0.5\textwidth]{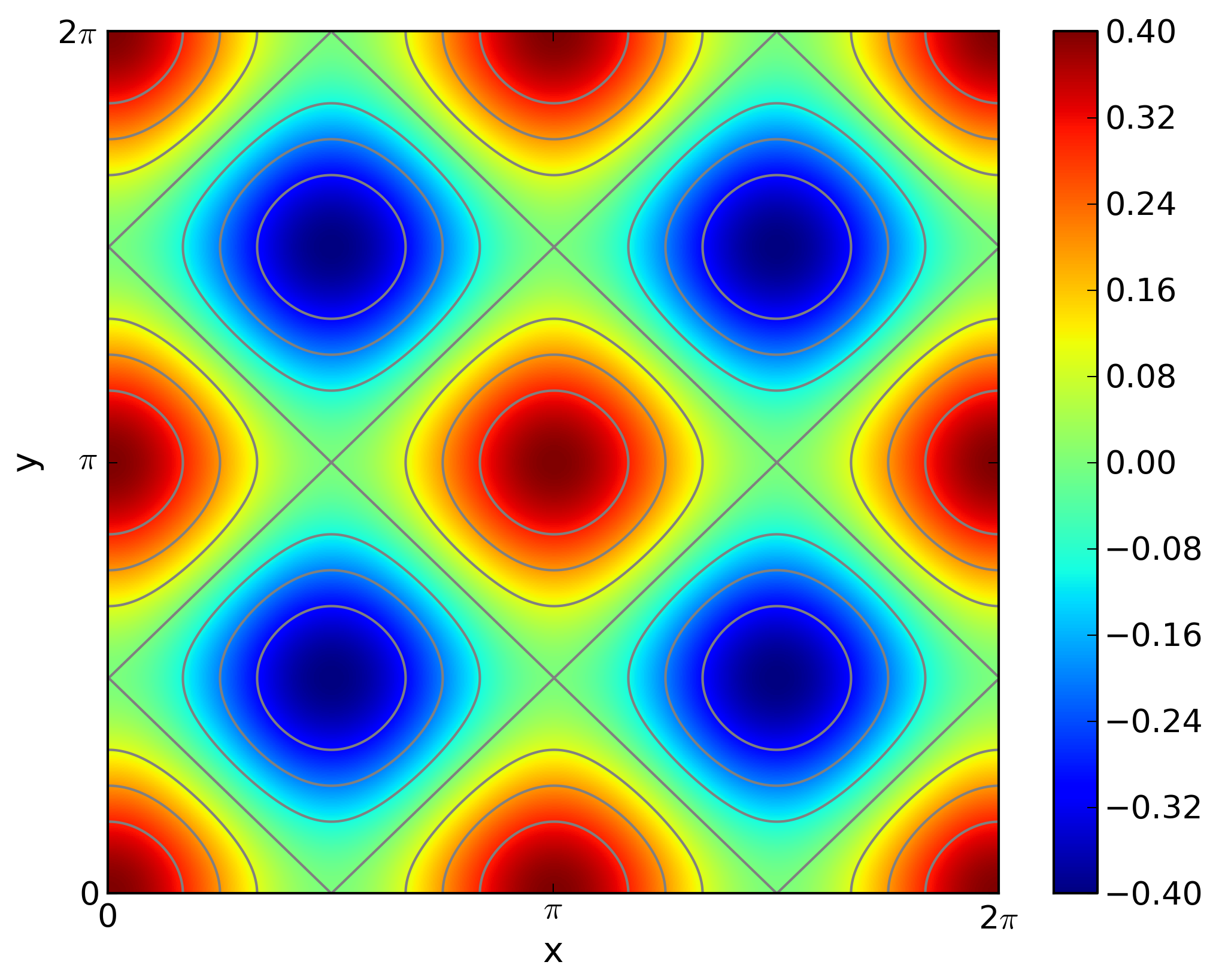}
\caption{Vorticity field of the laminar solution of~\eqref{eq:stokes}
in the absence of polymer feedback,
$\omega(\bm x)= -f_0 (\cos Kx + \cos Ky)/\nu K$,
for $K=2$, $f_0=0.02$ and $\nu=0.05$.}
\label{fig:cellular}
\end{figure}

In the absence of polymer feedback on the flow (i.e. $\mu=0$),
\eqref{eq:stokes} admits the fixed-point laminar solution: 
$\bm u=-\bm f/\nu K^2$.
The corresponding vorticity field $\omega(\bm x,t)$,
where $\omega\hat{\bm z}\equiv \bnabla\times\bm u$, is shown in 
figure~\ref{fig:cellular} and consists of a sequence of vortices
of alternate sign separated by lines of pure strain.
At time $t=0$, $\bm u$ is the fixed-point velocity field and
$\mathsfbi{C}$ is a perturbation of the identity matrix (we use
the same perturbation as in \citealt{TS09}, see their equation~(3)).
By using the length scale $1/K$ and the amplitude $f_0/\nu K^2$ of the
laminar velocity, we obtain the turnover time
$T=\nu K/f_0$. This allows us to define the Deborah number 
$\mathit{De}\equiv\tau/T$, which quantifies the ability of the flow to deform
polymers.

The reason for choosing the cellular force is twofold. On the one hand, 
it generates a flow structure in which the straining and vortical regions
are clearly separated, similarly to  the four-roll mill force considered in \cite{TS09} 
and \cite{TST11}---this feature will turn out useful in
highlighting the effect of artificial diffusivity on elastic turbulence.
On the other hand, the polymer stress
generated by the cellular force is less strong
than in the case of the four-roll mill force; thus
even in the absence of artificial diffusivity a moderate
spatial resolution is sufficient to compute the
velocity and polymer-stress fields accurately.

In order to preserve the positive definiteness of the conformation
tensor, we decompose $\mathsfbi{C}$ according to Cholesky, i.e. we
write $\mathsfbi{C}=\mathsfbi{L}\mathsfbi{L}^\top$, where
$\mathsfbi{L}$ is a lower triangular matrix and 
the diagonal elements $\mathsfi{L}_{ii}$ represent the eigenvalues of $\mathsfbi{C}$.
The positivity of $\mathsfbi{C}$ is then ensured by
evolving  $\ln\mathsfi{L}_{ii}$ instead of $\mathsfi{L}_{ii}$
\citep*{VC03,PMP06}.
The evolution equations for 
$\ln\mathsfi{L}_{ii}$ and $\mathsfi{L}_{ij}$ if $i\neq j$
\citep*[see][]{GPP15}
are solved on a regular grid with $1024^2$
collocation points by using a fourth-order finite difference
scheme for the spatial derivatives and a fourth-order Runge--Kutta 
scheme with time step $\mathrm{d}t=2\times 10^{-3}$ for the temporal integration.
The advection terms are treated according to the Kurganov--Tadmor
hyperbolic solver \citep{KT00}.
This solver was first applied to the numerical simulation of 
polymer solutions by \cite{VRBC06}. At locations where the original
finite-difference scheme would yield a too large gradient 
of $\mathsfi{L}_{ij}$ or $\ln\mathsfi{L}_{ii}$,
it selects a lower-order scheme that 
reduces the gradient.
Numerical schemes based on
the Kurganov--Tadmor solver have been employed in studies of both turbulent polymer
solutions (e.g. \citealt{VRBC07}; \citealt*{PMP10}; \citealt*{DVH10};
\citealt{RVCB10}; \citealt*{VdSP14}; \citealt{GPP15}; \citealt{graham18})
and elastic turbulence \citep{GP17,PGVG17}. 
Finally, the velocity field is obtained by solving the vorticity equation
associated with \eqref{eq:stokes} in Fourier space.

In order to assess the effect of artificial diffusivity, 
we also performed simulations which still use the Cholesky
decomposition and the Kurganov--Tadmor scheme as described above,
but in which we add the term
$\kappa\Delta\mathsfbi{C}$ to the right hand side
of \eqref{eq:conformation} (or rather the corresponding diffusion terms
to the equations for $\ln\mathsfi{L}_{ii}$ and $\mathsfi{L}_{ij}$).
We set $\kappa=5\times 10^{-5}$, so that
the Schmidt number $\mathit{Sc}\equiv\nu/\kappa=10^3$ is the same
as in previous numerical simulations of
elastic turbulence \citep{TS09,TST11,GCMB18}. 
This value of {\it Sc} is much higher than that
used in high-Reynolds-number simulations, where typically ${\it Sc}=0.5$ \citep[e.g.][]{G14},
but is nevertheless
three orders of magnitude smaller than it would be in reality
\citep{VRBC06}. However, it is not interesting to consider
values of $\it Sc$ much greater than $10^3$, because alone 
they are in general not sufficient to prevent numerical instabilities.

The parameters of the simulations are $K=2$, $f_0=0.02$, $\mu=0.01$, $\nu=0.05$,
$\tau=50$ and yield a Deborah number $\mathit{De}=10$. 
In particular,
the ratio $\mu/\nu$ is comparable to that used in previous simulations of
elastic turbulence. Additional simulations with a different choice
of parameters are reported in the online supplementary material and support
the results presented in the following section.

\section{Results}
\label{sect:results}

In this section, we compare numerical simulations of 
\eqref{eq:OB} based on the two approaches described in \S~\ref{sect:model},
i.e. using either  artificial diffusivity with
$\mathit{Sc}=10^3$ or the Kurganov--Tadmor scheme, for which 
$\mathit{Sc}=\infty$.
As we shall see, the effect of artificial diffusivity is so big
that to describe it,
it is sufficient to examine the qualitative properties of the flow.

\begin{figure}
\centering
\includegraphics[width=0.49\textwidth]{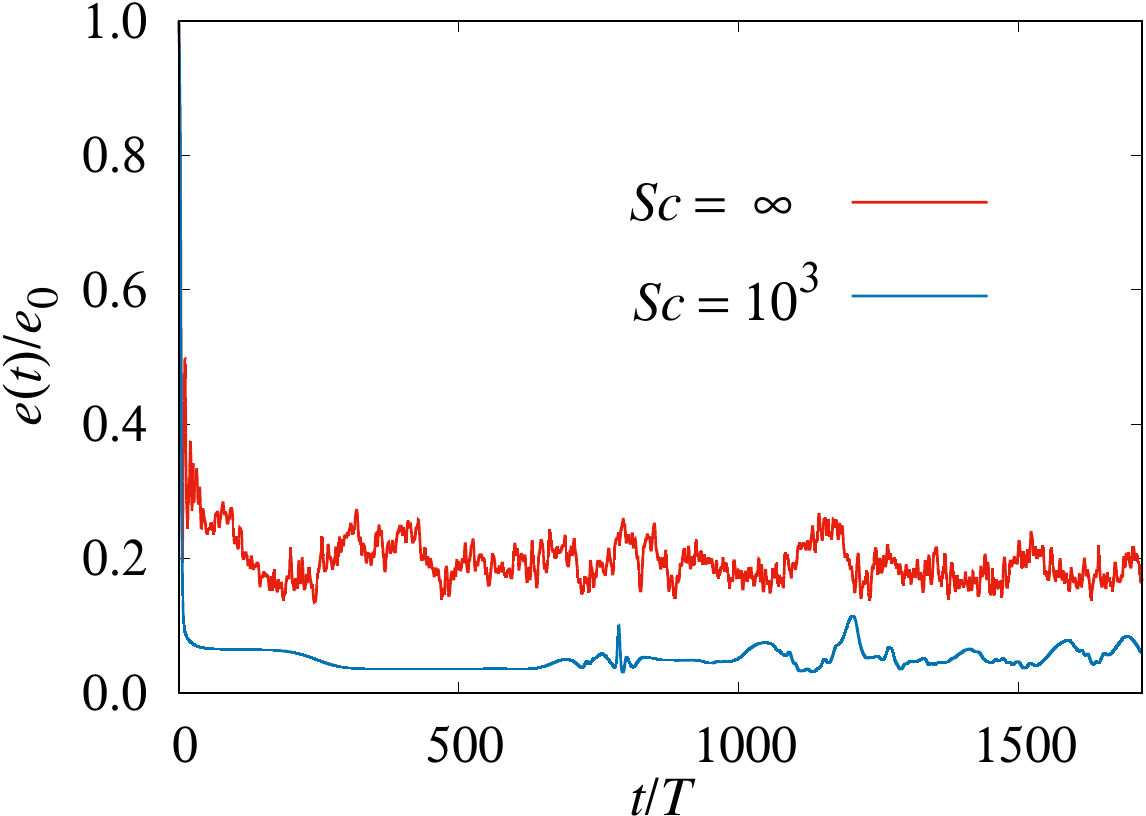}\hfill%
\includegraphics[width=0.5\textwidth]{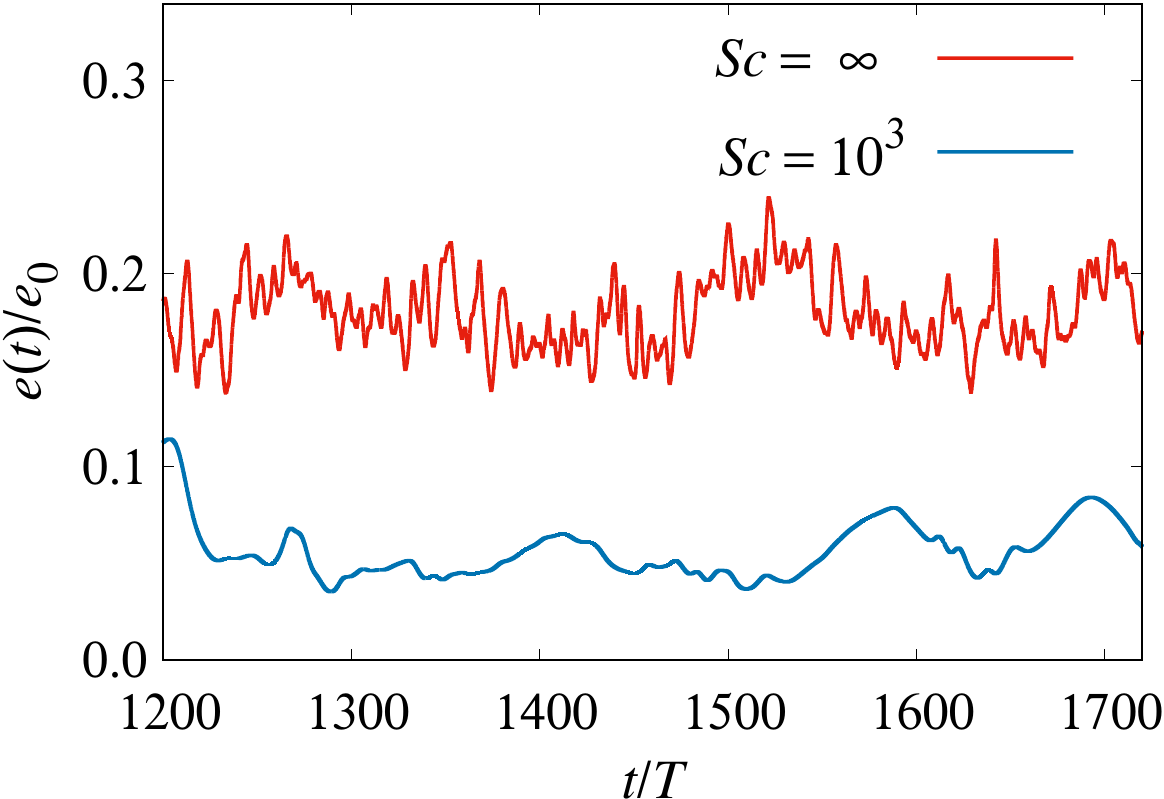}
\\[3mm]
\caption{Left: Time series of the kinetic energy of the polymer solution
rescaled by the kinetic energy of the fixed-point laminar flow ($\mu=0$),
i.e. $e_0\equiv f_0^2/2\nu^2 K^4$, for ${\it Sc}=\infty$ (red, top curve)
and for ${\it Sc}=10^3$ (blue, bottom curve).
Right: zoom of the left panel in the steady state.}
\label{fig:kinetic}
\end{figure}

Figure~\ref{fig:kinetic} shows the time series of the kinetic energy 
$e(t)\equiv\frac{1}{2}\int_V \vert\bm u(\bm x,t)\vert^2\mathrm{d}\bm x$.
For ${\it Sc}=10^3$, the system remains in an almost frozen 
state for a long time,
after which it becomes chaotic.
If the simulation is not long enough,
such an initial frozen state may lead to a wrong
interpretation of the dynamics, since 
the flow regime may be erroneously described as laminar.
The behaviour of $e(t)$ is analogous to that found in \citet{TS09} and \cite{TST11}
for a four-roll mill force and same value of $\it Sc$. 
For ${\it Sc}=\infty$, in contrast, the motion of the 
solution becomes chaotic much more rapidly and, in the steady state, 
the kinetic energy fluctuates at a frequency much higher than
when artificial diffusivity is present. Moreover, the mean kinetic energy 
is greater.

\begin{figure}
\centering
\includegraphics[width=0.49\textwidth]{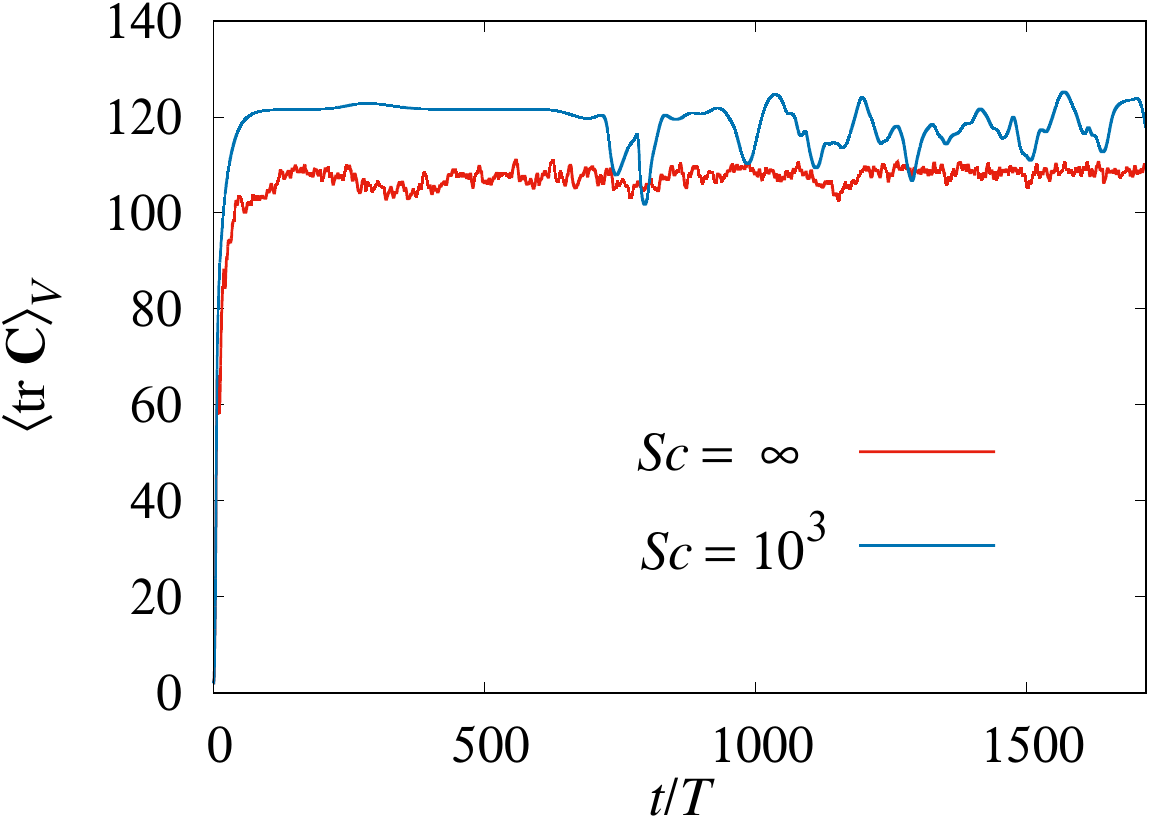}\hfill%
\includegraphics[width=0.5\textwidth]{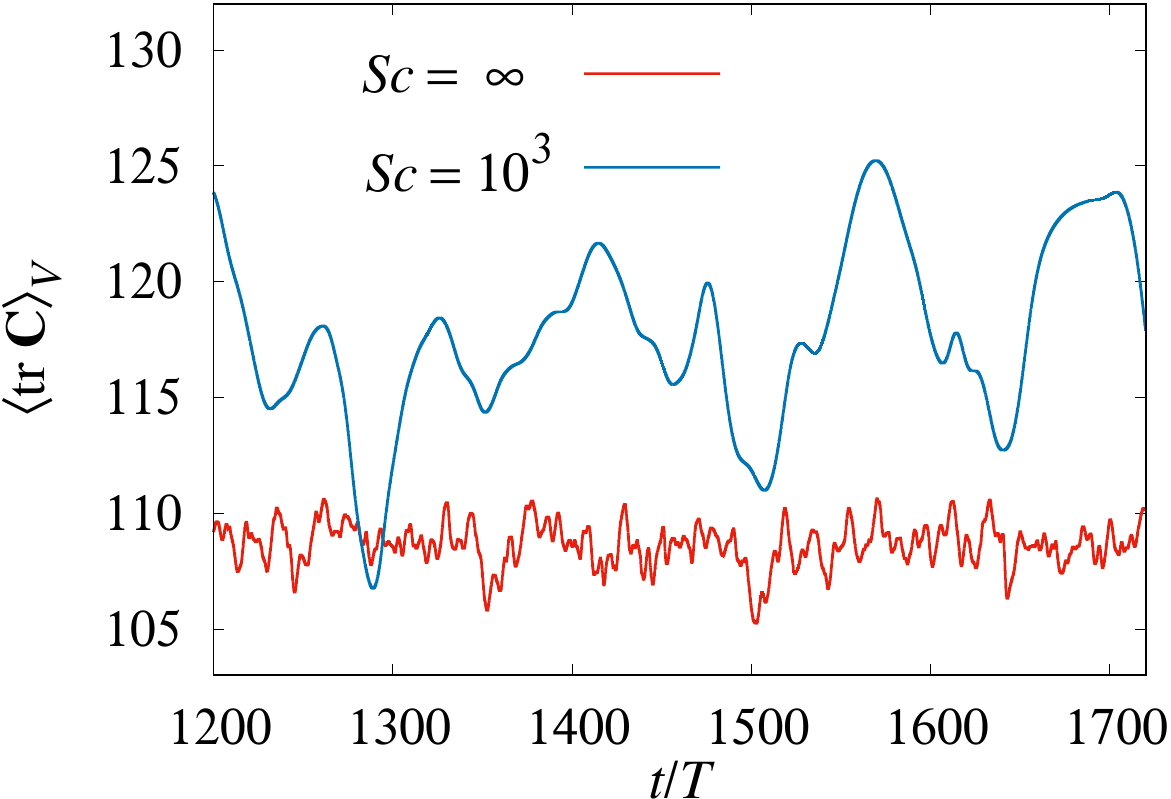}
\\[3mm]
\caption{Left: Time series of the trace of the polymer conformation tensor
averaged over the spatial domain~$V$
for ${\it Sc}=10^3$ (blue, top curve)
and ${\it Sc}=\infty$ (red, bottom curve).
Right: zoom of the left panel in the steady state.}
\label{fig:rp2}
\end{figure}

The time series
of the trace of $\mathsfbi{C}$ averaged over $V$
show behaviours analogous
to those of $e(t)$, with $\langle\operatorname{tr}\mathsfbi{C}\rangle_V$
displaying much slower oscillations when $\mathit{Sc}=10^3$ and
this only after a long, initial, almost frozen state (figure~\ref{fig:rp2}).


\begin{figure}
\includegraphics[width=0.5\textwidth]{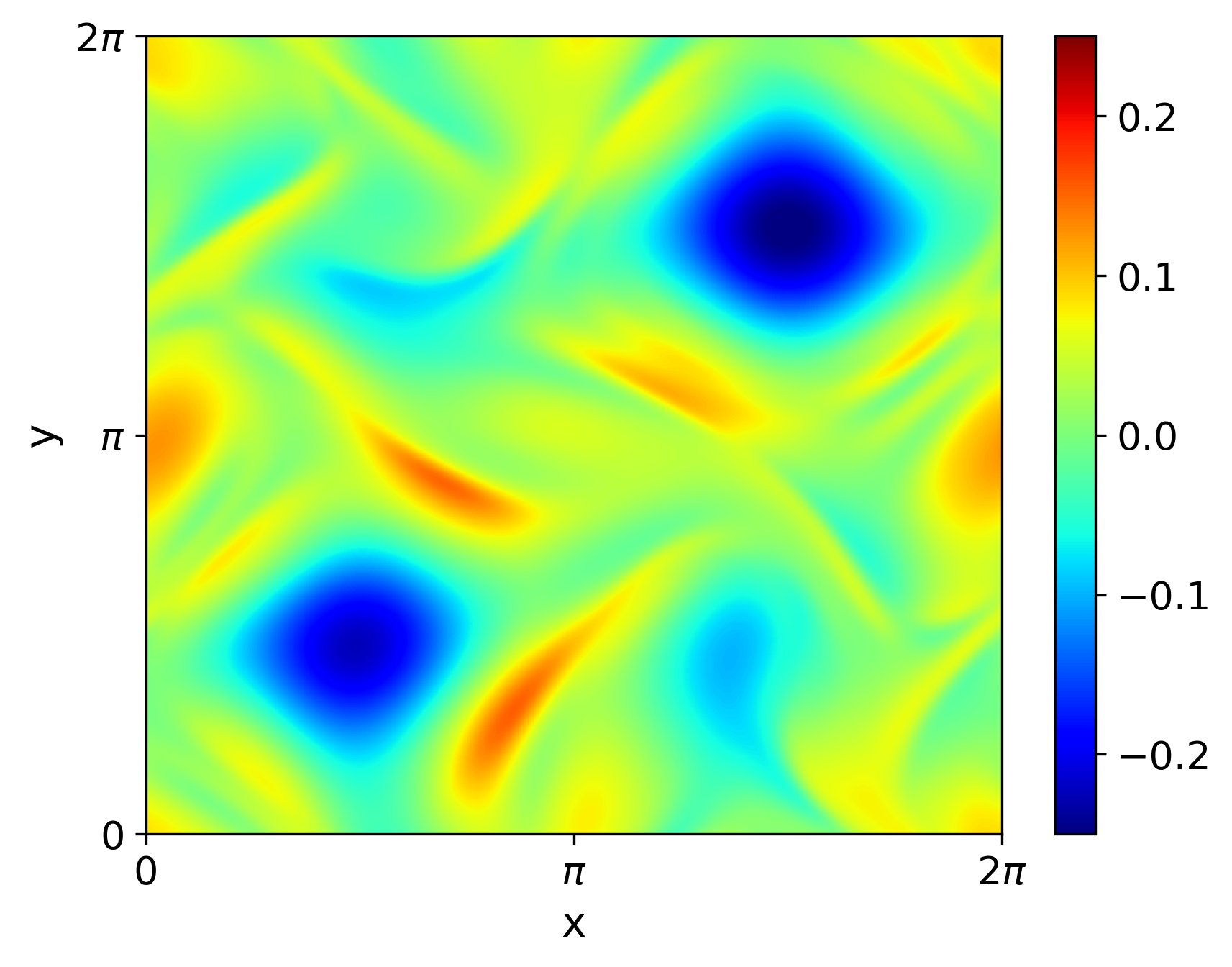}\hfill%
\includegraphics[width=0.5\textwidth]{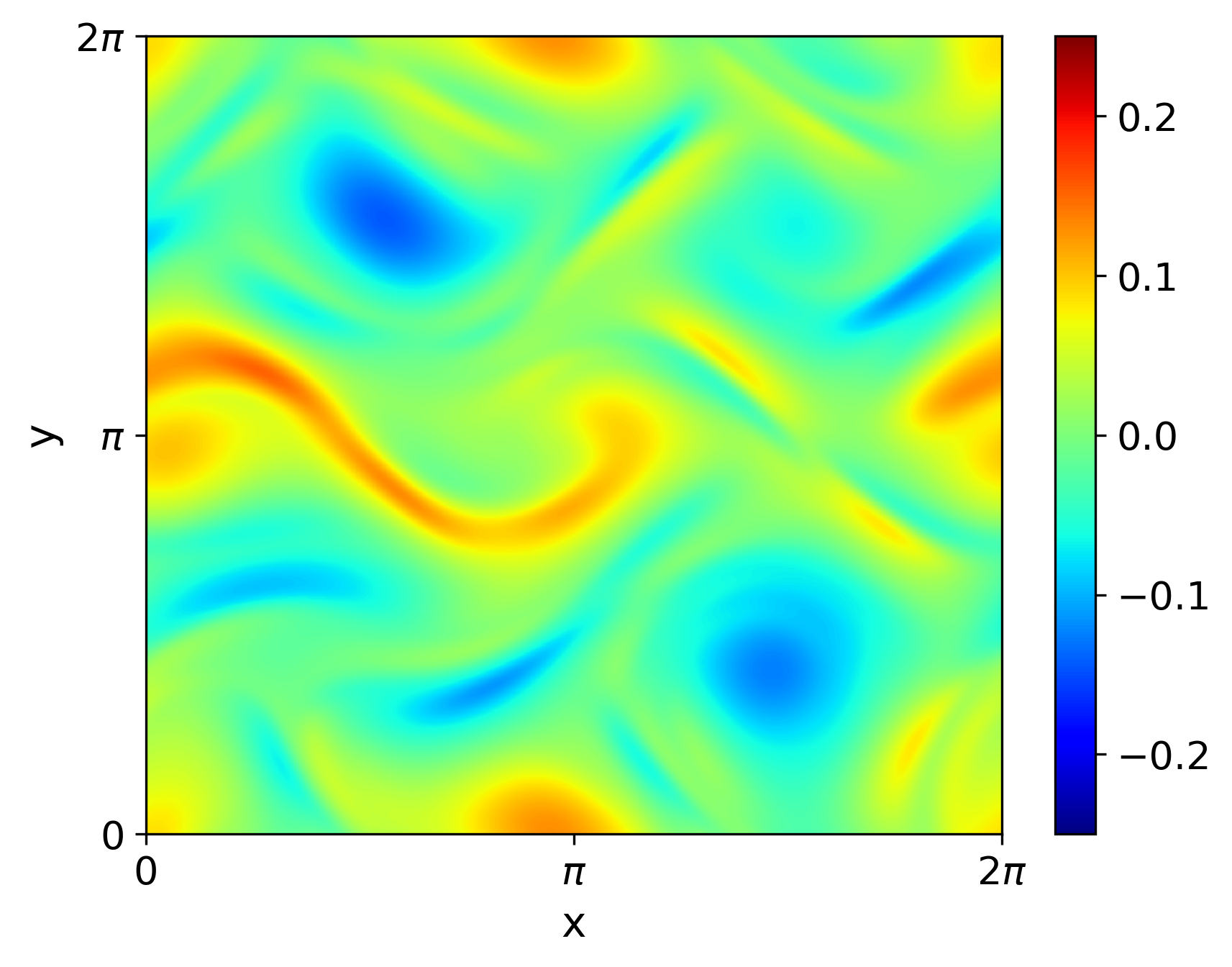}%
\caption{Snapshot of the vorticity field for ${\it Sc}=10^3$
at $t/T=1200$ (left) and $t/T=2212$ (right).}
\label{fig:vorticity-sc}
\end{figure}

\begin{figure}
\includegraphics[width=0.5\textwidth]{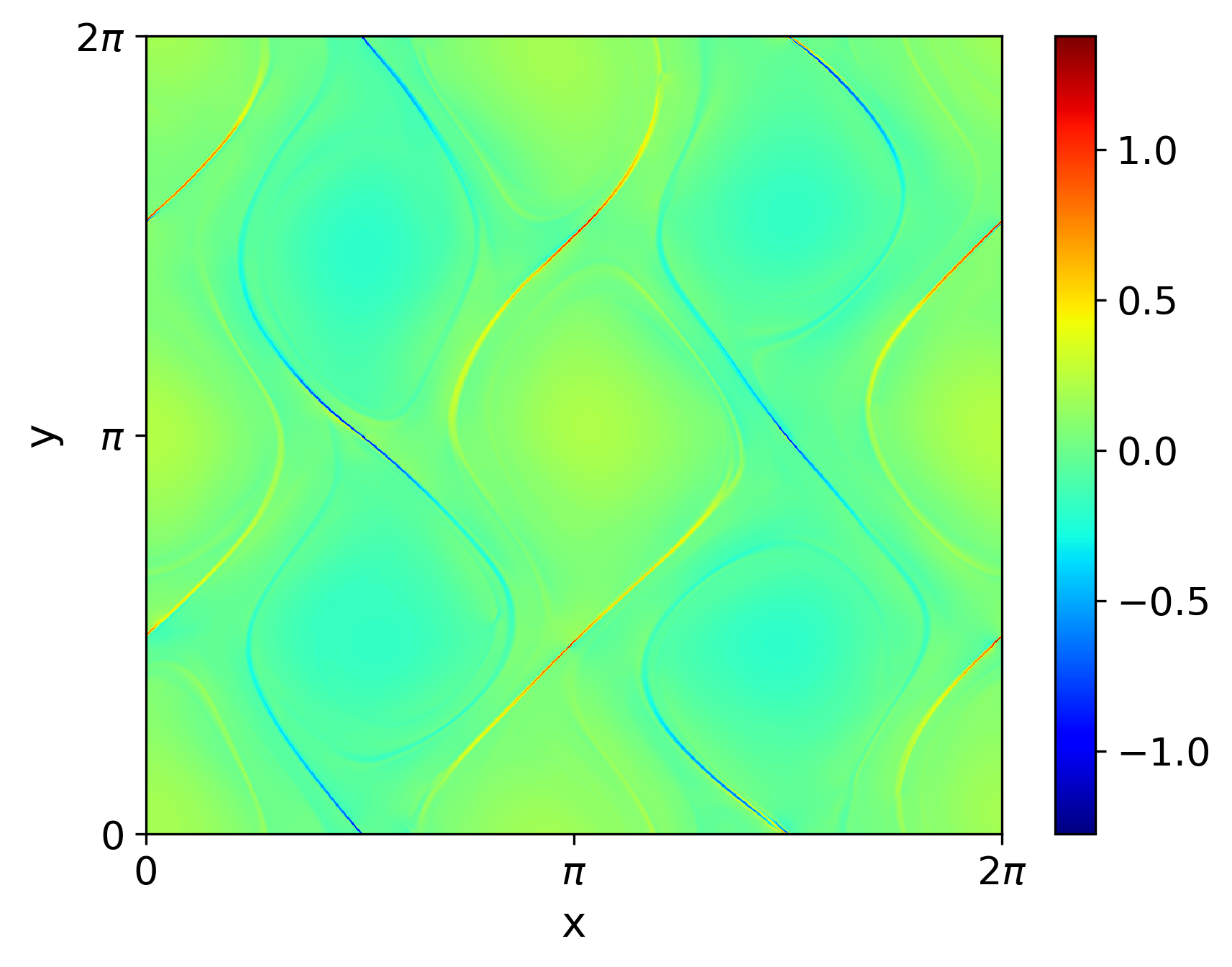}\hfill%
\includegraphics[width=0.5\textwidth]{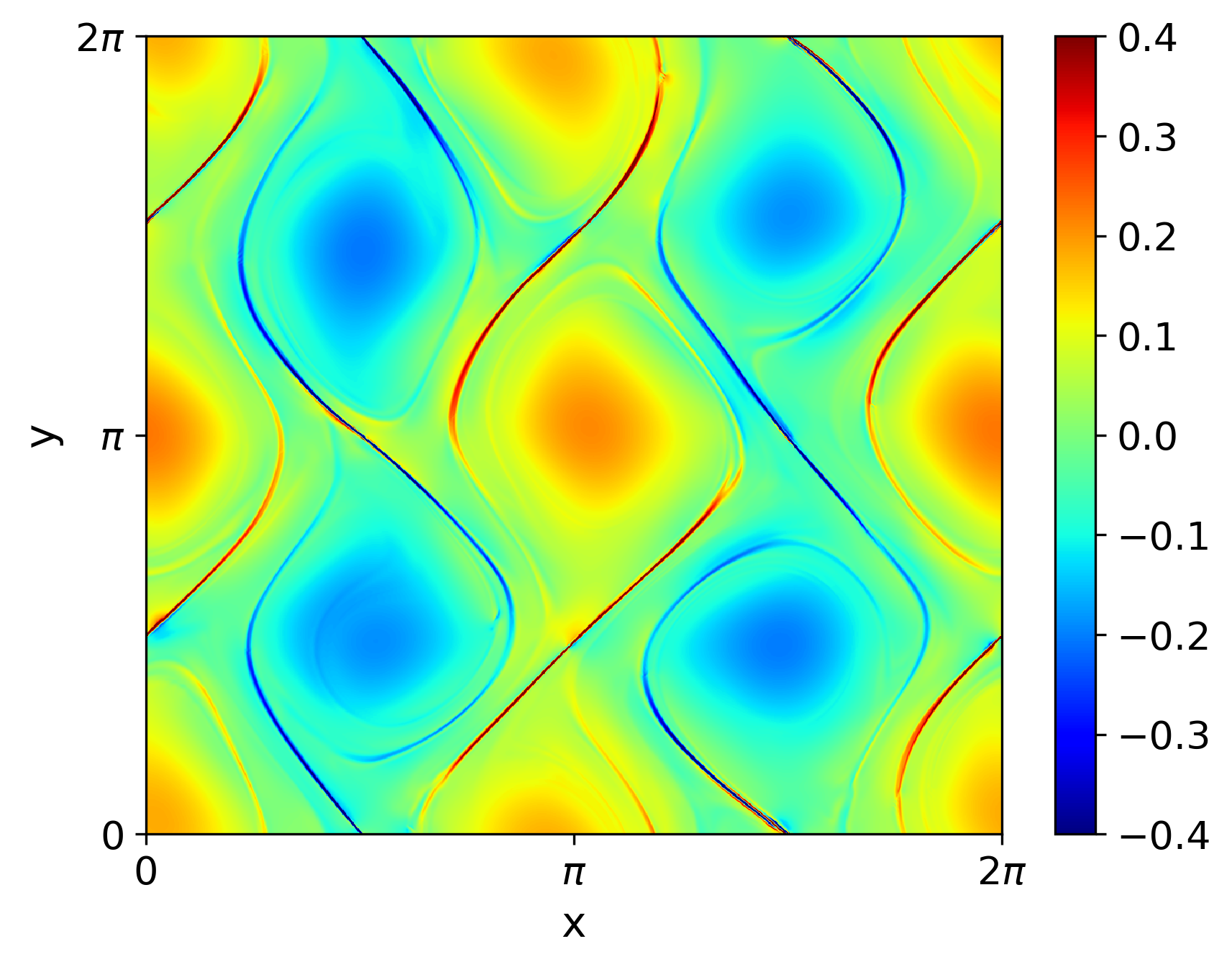}%
\caption{Left: 
a representative snapshot of the vorticity field for ${\it Sc}=\infty$.
Right: the same snapshot as in the left panel but with a
rescaled colour bar. The purpose of this rescaling is to show the cellular
structure of the flow more clearly.}
\label{fig:vorticity-kt}
\end{figure}

The snaphots of the vorticity field also indicate a striking
difference between the two integration methods (figures~\ref{fig:vorticity-sc}
and~\ref{fig:vorticity-kt}).
With  artificial diffusivity, 
once the system starts
fluctuating the spatial structure of the flow departs from
that which would be imposed by the force (figure~\ref{fig:vorticity-sc}). 
Only some of the 
vortical cells continue to exist, whilst others break down
and patches of vorticity contaminate the cellular 
structure of the base flow. Furthermore, 
the number and the location of unbroken
vortical cells vary in time (see the snapshots at two different times
in figure~\ref{fig:vorticity-sc}).
This dynamics is equivalent to the symmetry-breaking transition
observed by \cite{TS09} and \cite{TST11}.
Contrastingly, for ${\it Sc}=\infty$ 
the vorticity field displays fluctuations which perturb the cellular
vortices, but its
large-scale structure essentially remains slaved to that of the background force
(see figure~\ref{fig:vorticity-kt} for a representative snapshot).
Large perturbations of the 
vorticity field are concentrated on thin filaments located
in the vicinity of the lines of pure strain which separate the
vortical cells.
Thus, the symmetry breaking shown in figure~\ref{fig:vorticity-sc} is a
spurious effect due to artificial diffusivity.


\begin{figure}
\centering
\includegraphics[width=0.5\textwidth]{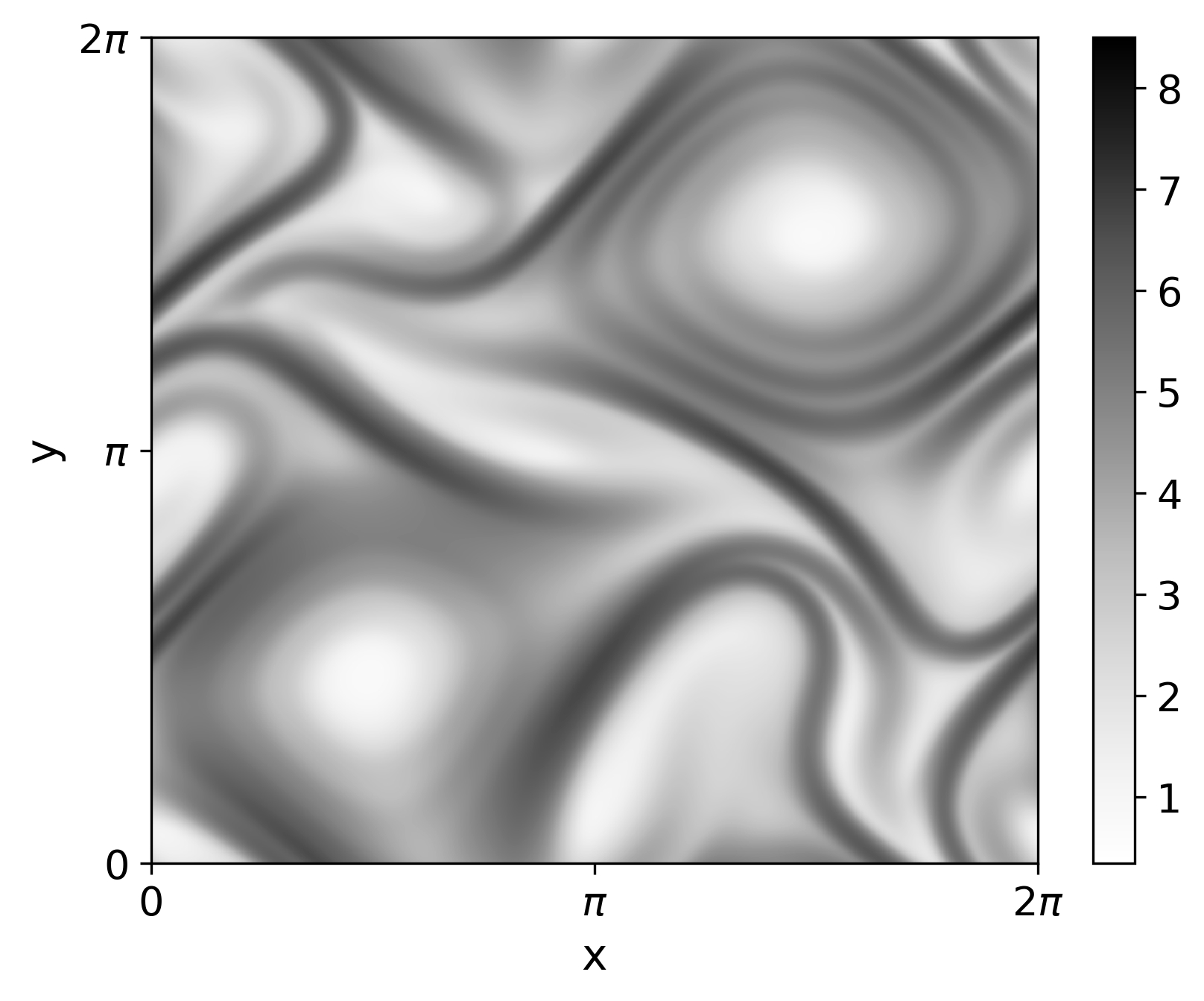}%
\hfill%
\includegraphics[width=0.5\textwidth]{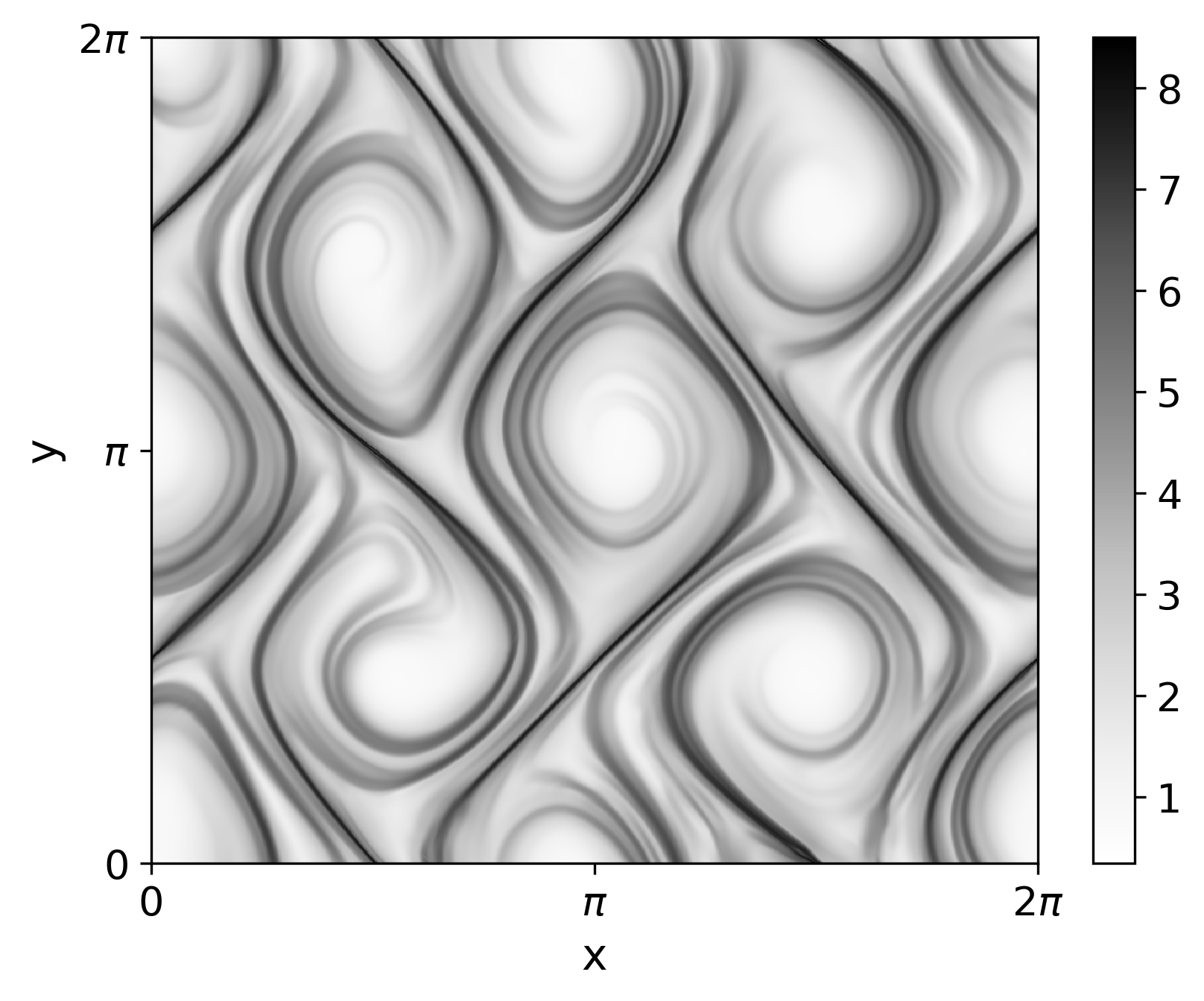}
\caption{Snapshots of $\ln(\operatorname{tr}\mathsfbi{C})$
for ${\it Sc}=10^3$ (left) and  ${\it Sc}=\infty$ (right).
The snapshots are taken at the same times as in the left panel
of figure~\ref{fig:vorticity-sc} and in
figure~\ref{fig:vorticity-kt}, respectively.
}
\label{fig:stress}
\end{figure}

The snapshots of $\operatorname{tr}\mathsfbi{C}$ also
show a strong qualitative difference in the behaviour of 
the polymer stress (figures~\ref{fig:stress}). 
The analysis of these snapshots allows us to understand the
effect of  artificial diffusivity.
Rather than the Schmidt number, 
here the relevant dimensionless parameter is the 
P\'eclet number ${\it Pe}=f_0/\kappa\nu K^3$, 
which is the ratio
of the time scale associated with polymer-stress diffusion, 
$(\kappa K^2)^{-1}$,
 and the time scale at which the polymer stress is convected, $T$.
When $\mathit{Sc}=\infty$, $\mathit{Pe}$ is also infinite and
the evolution of the polymer stress
is dominated by convection. In this case,
highly-stretched polymers---and hence large polymer stresses---are 
essentially found in strain-dominated regions of
the flow, while the polymer stress is weak in vortical regions, where polymers
rapidly contract. 
Thus, in the cellular flow, the vorticity field is strongly 
affected by polymer stresses at the boundaries of the cells; inside
the cells, the external force dominates and imposes the cellular structure.
However, when polymer-stress diffusion becomes relevant 
(i.e. $\mathit{Sc}=10^3$ and, with our choice of parameters, $\mathit{Pe}=10^3$), 
large polymer stresses spread
far from the straining lines 
where they are created and reach the interior of 
the vortices; this 
destabilizes the cellular structure and generates the 
symmetry breaking observed at ${\it Sc}=10^3$.
Furthermore, we note that 
although for ${\it Sc}=\infty$ polymers can locally be 
highly stretched (figure~\ref{fig:stress}), the average polymer
stretching is generally higher for ${\it Sc}=10^3$ than for
${\it Sc}=\infty$ (figure~\ref{fig:rp2}).

\cite{zaki18} recently proposed to quantify the deviation of the polymer
configuration from the equilibrium one by considering
the geodesic distance
between $\mathsfbi{C}$ and $\mathsfbi{I}$ in the space of positive-definite
tensors. The snapshots of this distance (see the supplementary material) 
confirm the behaviours observed
in the snapshots of $\operatorname{tr}\mathsfbi{C}$ 
(figure~\ref{fig:stress}).

\begin{figure}
\centering
\includegraphics[width=0.5\textwidth]{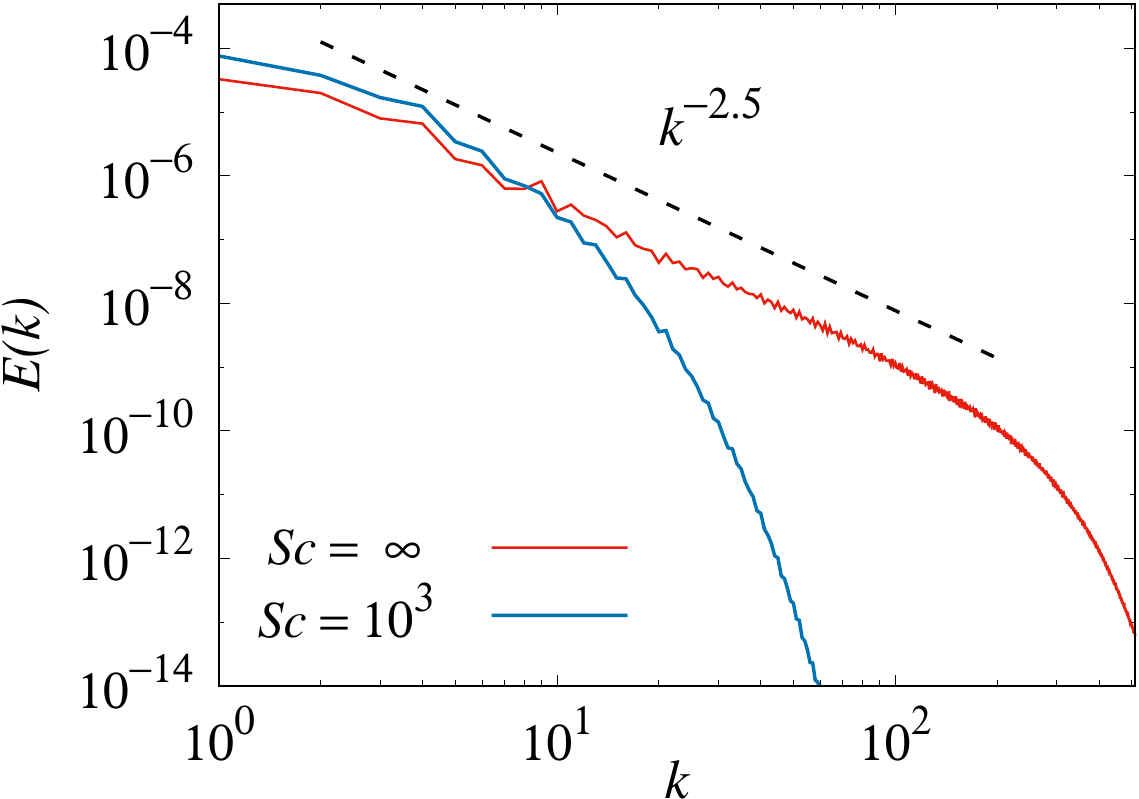}%
\hfill%
\includegraphics[width=0.49\textwidth]{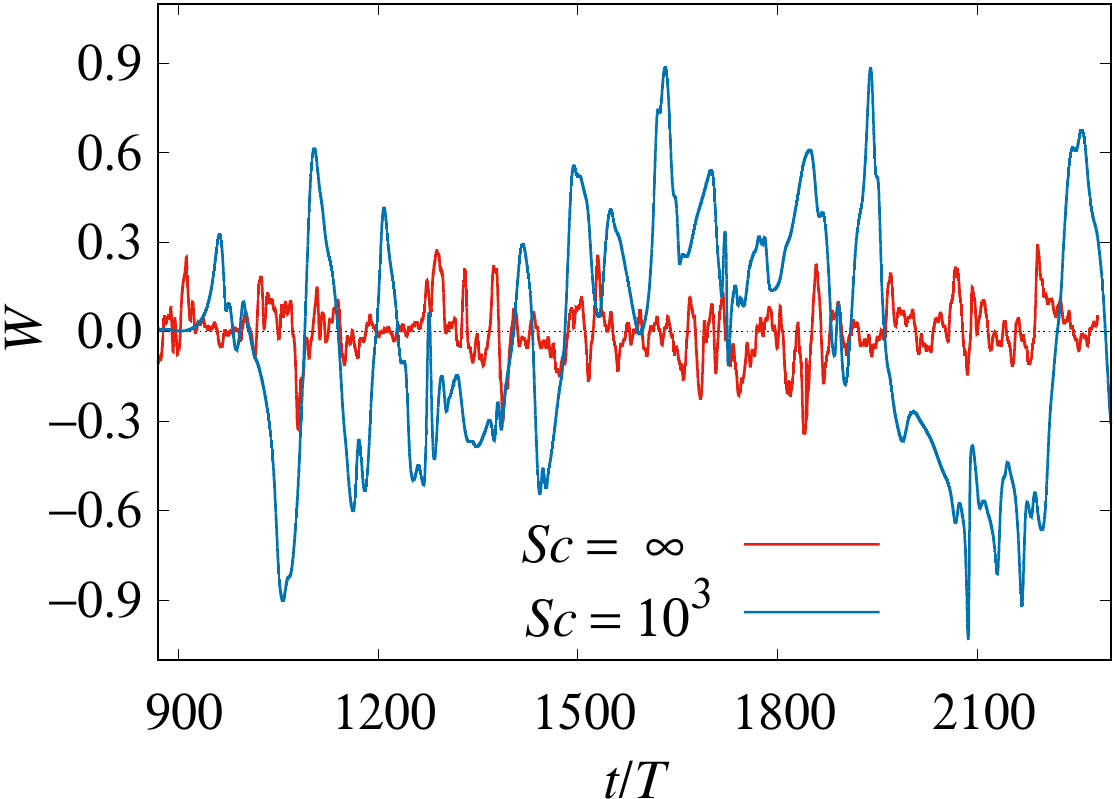}
\caption{Left: Kinetic-energy spectrum for 
${\it Sc}=10^3$ and ${\it Sc}=\infty$.
For ${\it Sc}=\infty$ the spectrum behaves as the power-law $k^{-2.5}$.
Right: Time series of $W$. 
For ${\it Sc}=10^3$ , $W$ displays much larger fluctuations 
than for ${\it Sc}=\infty$.}
\label{fig:spectra}
\end{figure}

Figures~\ref{fig:vorticity-sc} and~\ref{fig:vorticity-kt}
also indicate that the vorticity field is
much smoother
for ${\it Sc}=10^3$ than for ${\it Sc}=\infty$. 
This can be understood by noting that,
in elastic turbulence, the small scales 
of the flow
are not activated by a cascade phenomenon but rather by
the fluctuations of the polymer feedback at the same scales,
and for $\mathit{Sc}=10^3$
the high-wave-number fluctuations of the polymer stress
are damped by diffusivity. 
%
%
%
The smoothness of the flow can be quantified
by examining the spectrum of the velocity fluctuations,
$E(k)\equiv\sum_{k-1/2< k' \le k+1/2} \langle |\hat {\bm v}({\bm k'},t)|^2 \rangle _t$,
where 
$\langle\bcdot\rangle_t$ denotes the time average over the steady state
and $\hat{\bm v}(\bm k,t)$ is the Fourier transform of the velocity field 
minus its time average.
We find again that artificial diffusivity has a strong impact on the flow
(figure~\ref{fig:spectra}, left).
For ${\it Sc}=10^3$, $E(k)$ decay very rapidly with the wave number,
whereas for ${\it Sc}=\infty$ it behaves as a power law:
$E(k)\propto k^{-2.5}$.
Thus, fluctuations are sustained at much smaller scales when ${\it Sc}=\infty$.
The power law is shallower than those found previously in experiments and
numerical simulations with different forcings, in which the exponent
varied with the setup but
was always smaller than $-3$
\citep{GS00,GS04,BBBCM08,RV16,WG13,WG14,GP17}.
A kinetic-energy spectrum steeper than $k^{-3}$ was also predicted by
\cite{FL03}. 
This prediction, however, does not necessarily 
apply to the case under consideration,
because it assumes statistical homogeneity and isotropy and our flow
does not enjoy these statistical symmetries.
(Note that, for the Kolmogorov force,
the same integration
scheme used here yields the exponent $-3.7$ \citep{PGVG17}, 
which agrees with the results
of \citet{BBBCM08}; see also \cite{BB10} and \cite{GCMB18}.) 

To quantify the breaking of the cellular symmetry, we also consider the ratio
\begin{equation}
W=\dfrac{\omega(0,0)-\omega(\upi,0)}{\omega(0,0)+\omega(\upi,0)},
\end{equation}
which compares the amplitute of vorticity at two positions in the flow,
$(x,y)=(0,0)$ and $(x,y)=(\upi,0)$.
For the laminar solution ($\mu=0$), these two positions are the 
centres of two equal-sign vortices and $W=0$. Nonzero values
of $W$ therefore correspond to deviations from the cellular symmetry.
Figure~\ref{fig:spectra} (right)
shows that the fluctuations of $W$ are much bigger for
${\it Sc}=10^3$ than for ${\it Sc}=\infty$
(analogous results are found for other couples of equal-sign vortices).
This confirms that the cellular structure is strongly modified by
artificial diffusivity.

\section{Conclusions}
\label{sect:conclusions}

In numerical simulations of constitutive models of polymer solutions
the addition of  artificial diffusivity
is a well-known strategy for overcoming the numerical 
instabilities generated by the loss of positive definiteness of the 
polymer-stress tensor.
We have studied the accuracy of this approach
in elastic turbulence
by comparing simulations of the two-dimensional Oldroyd-B model
sustained by a cellular external
force, with and without  artificial diffusivity.
Our results show that  artificial diffusivity
has a strong impact on the qualitative spatial and temporal properties
of the flow. In particular, with  artificial diffusivity
the cellular structure imposed by the external force breaks down,
as also found by \cite{TS09} and \cite{TST11}
for a four-roll mill force and the same value of ${\it Sc}$.
Without  artificial diffusivity, 
the flow is mainly perturbed in strain-dominated regions but,
albeit chaotic,
essentially reproduces the symmetries of the background force.

We have also performed simulations of the FENE-P model in exactly the
same setting and found similar results (see Appendix~\ref{appendix});
hence the effect of  artificial diffusivity is not modified by
the nonlinearity of the elastic force.

To preserve the positive definiteness of the polymer
conformation tensor and accurately resolve its gradient,
we have used a combination of the Cholesky decomposition and 
the Kurganov--Tadmor hyperbolic solver. 
Adding artificial diffusivity to this scheme produces
the migration of high polymer stresses to regions of the flow
where in principle polymers would be weakly stretched. This
generates strong differences in the large-scale flow.
We believe that this effect of artificial diffusivity 
is not specific to the Kurganov--Tadmor scheme, i.e.
the addition of an excessively large polymer-stress diffusion is
expected to produce analogous modifications of the large-scale flow
also in other integration schemes and, vice versa, if 
a scheme does not use artificial diffusivity
or similar dissipation terms, it should not produce
the symmetry breaking discussed above. 

We have described the motion of the fluid by means of the 
Stokes equations, as in some of the previous studies on elastic turbulence.
Other studies have used the Navier--Stokes equations.
Thus, we have also performed numerical simulations of \eqref{eq:conformation}
coupled with the Navier--Stokes equations with the same cellular force.
The Reynolds number ${\it Re}=f_0/\nu^2 K^3$ has been set to unity,
which is below the critical value for the appearance of inertial instabilities, 
${\it Re}_c=\sqrt{2}$ \citep{GY84}. The conclusions on
the effect of  artificial diffusivity
are exactly the same as when the Stokes equations are used.

In our study,
we specifically selected the cellular force in order
to illustrate the impact of 
artificial diffusivity on numerical simulations of elastic turbulence.
This force indeed generates a flow in which
the straining and vortical regions are distinct. Hence in the absence
of  artificial diffusivity
the polymer stress is chiefly located at the straining regions,
whereas in the presence of diffusivity it spreads outside them.
For other external forces that mix strain and vorticity, the effect of
 artificial diffusivity
may be less dramatic. This is the case, for instance, of the Kolmogorov 
force, which in the absence of polymer feedback and at low Reynolds numbers
generates a sinusoidal shear. For the Kolmogorov force, the velocity spectrum
indeed 
behaves as a power law even for ${\it Sc}=10^3$ \citep{GCMB18}, although
of course the power law is observed on a smaller range of 
wave numbers compared to simulations that do not use
 artificial diffusivity \citep{PGVG17}.

The effect of artificial diffusivity is more dramatic in elastic
turbulence than at high Reynolds numbers or in laminar flows. 
In these latter cases, indeed,
artificial diffusivity produces quantitative changes in the solution
of the Oldroyd-B and FENE-P models but
does not modify the qualitative 
behaviour of the velocity and polymer-conformation fields, provided
of course {\it Sc} is not excessively small 
(see, e.g., figures~8 to~10 in \citealt{VRBC06}
and figures~3 and~4 in \citealt{STD18} for the high-Reynolds-number regime
and figure~7 in \citealt{T11} for the laminar regime).
In elastic turbulence, before the addition of polymers the flow is
laminar.
Artificial diffusivity thus strongly
modifies the large-scale flow by bringing high polymer stresses
to regions of the fluid where the polymers would be weakly
stretched and hence the flow would be weakly chaotic.
This effect is peculiar to the regime of elastic turbulence
and can lead to the spurious behaviours described above. Therefore,
great caution should be taken in using 
artificial diffusivity to prevent numerical instabilities in simulations
of elastic turbulence.



\acknowledgments
We would like to thank T. Matsumoto for useful discussions.
The computations were performed at Centre de Calcul Interactif, Universit\'e Nice Sophia Antipolis
and M\'esocentre SIGAMM, Observatoire de la C\^ote d'Azur.

\appendix

\section{Effect of artificial diffusivity in the FENE-P model}
\label{appendix}

A drawback of the Oldroyd-B model is that there is not
a maximum polymer extension.
Infinitely large extensions are
in principle allowed, and in certain flow conditions
$\operatorname{tr}\mathsfbi{C}$ grows indefinitely.
In the FENE-P model,  a maximum polymer extension
$\ell_{\rm max}$ is imposed by modifying \eqref{eq:OB} as follows
\begin{subequations}
\begin{equation}
\bnabla p=\nu \Delta\bm u +\dfrac{\mu}{\tau}\,\bnabla\bcdot[h(\operatorname{tr}\mathsfbi{C})
\mathsfbi{C}]+\bm f,
\qquad \bnabla\bcdot\bm u=0,
\end{equation}
\begin{equation}
\partial_t \mathsfbi{C}+\bm u\bcdot\bnabla\mathsfbi{C}=(\bnabla\bm u)\bcdot\mathsfbi{C}
+\mathsfbi{C}\bcdot(\bnabla\bm u)^\top
-\dfrac{h(\operatorname{tr}\mathsfbi{C})\mathsfbi{C}-\mathsfbi{I}}{\tau},%
\label{eq:C-FENE-P}
\end{equation}%
\label{eq:FENE-P}
\end{subequations}
where
\begin{equation}
h(\zeta)=\dfrac{\ell^2_{\rm max}-2}{\ell^2_{\rm max}-\zeta}.
\end{equation}
The function $h(\operatorname{tr}\mathsfbi{C})$ diverges as $\operatorname{tr}\mathsfbi{C}$
approaches $\ell_{\rm max}^2$ and hence causes $\operatorname{tr}\mathsfbi{C}$
to stay smaller than $\ell_{\rm max}^2$. 

\begin{figure}
\centering
\includegraphics[width=0.51\textwidth]{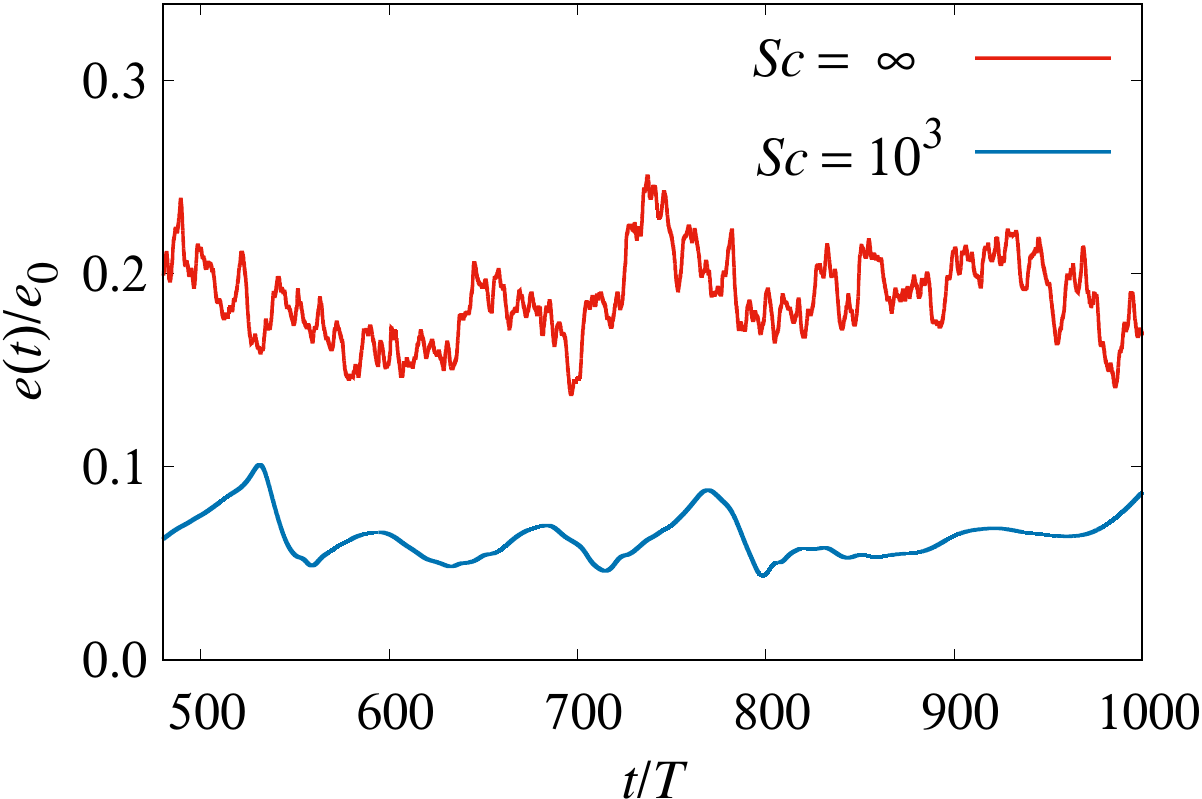}\hfill%
\includegraphics[width=0.485\textwidth]{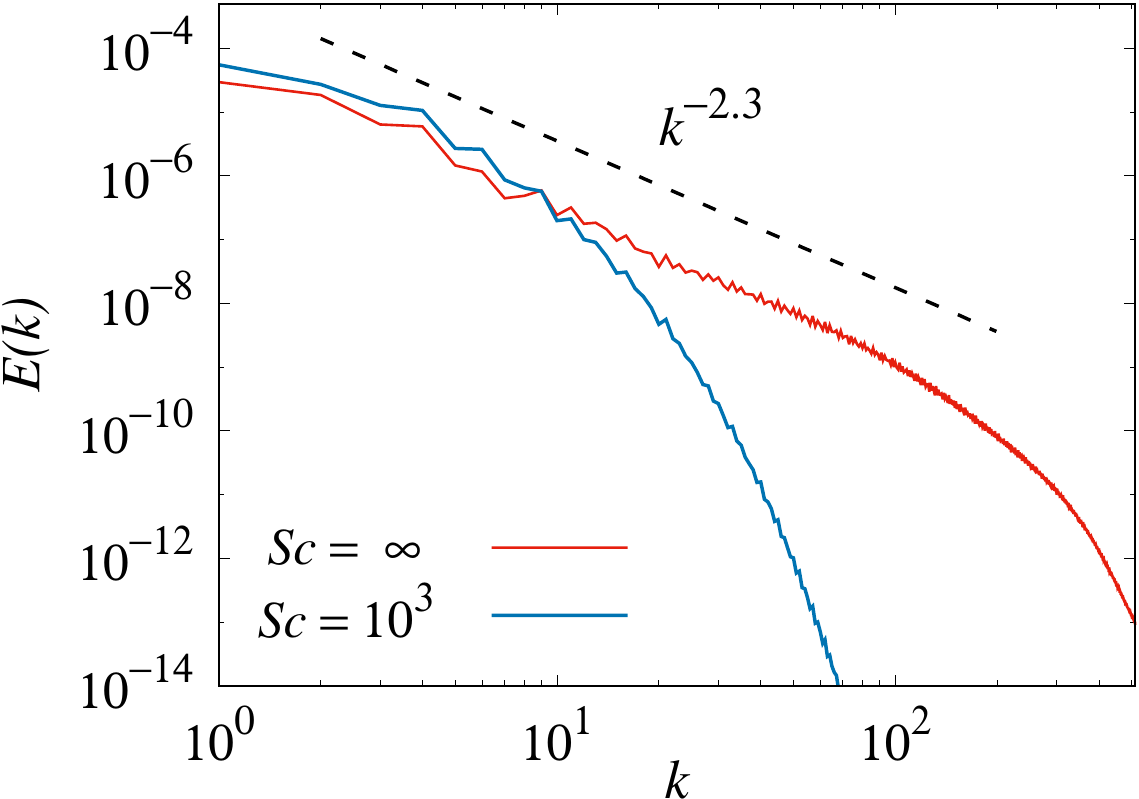}%
\caption{FENE-P model. Left:
Time series of the rescaled kinetic energy 
in the steady state for ${\it Sc}=\infty$ (red, top curve) and
${\it Sc}=10^3$ (blue, bottom curve).
Right:
Kinetic-energy spectra
for 
${\it Sc}=10^3$ and ${\it Sc}=\infty$.
For ${\it Sc}=\infty$ the spectrum behaves as the power-law $k^{-2.3}$.
}
\label{fig:spectra-fene-p}
\end{figure}

\begin{figure}
\centering
\includegraphics[width=0.5\textwidth]{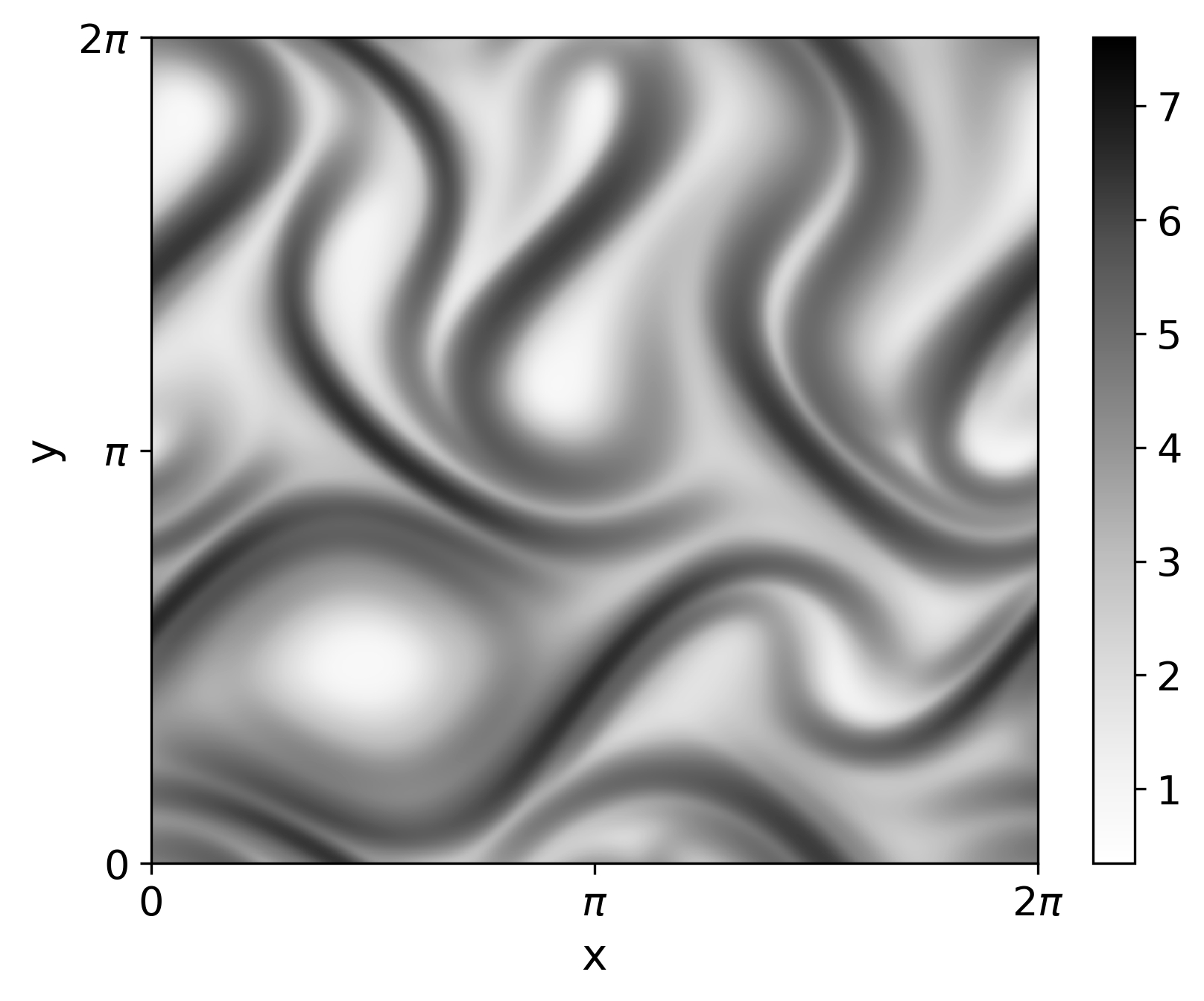}%
\hfill%
\includegraphics[width=0.5\textwidth]{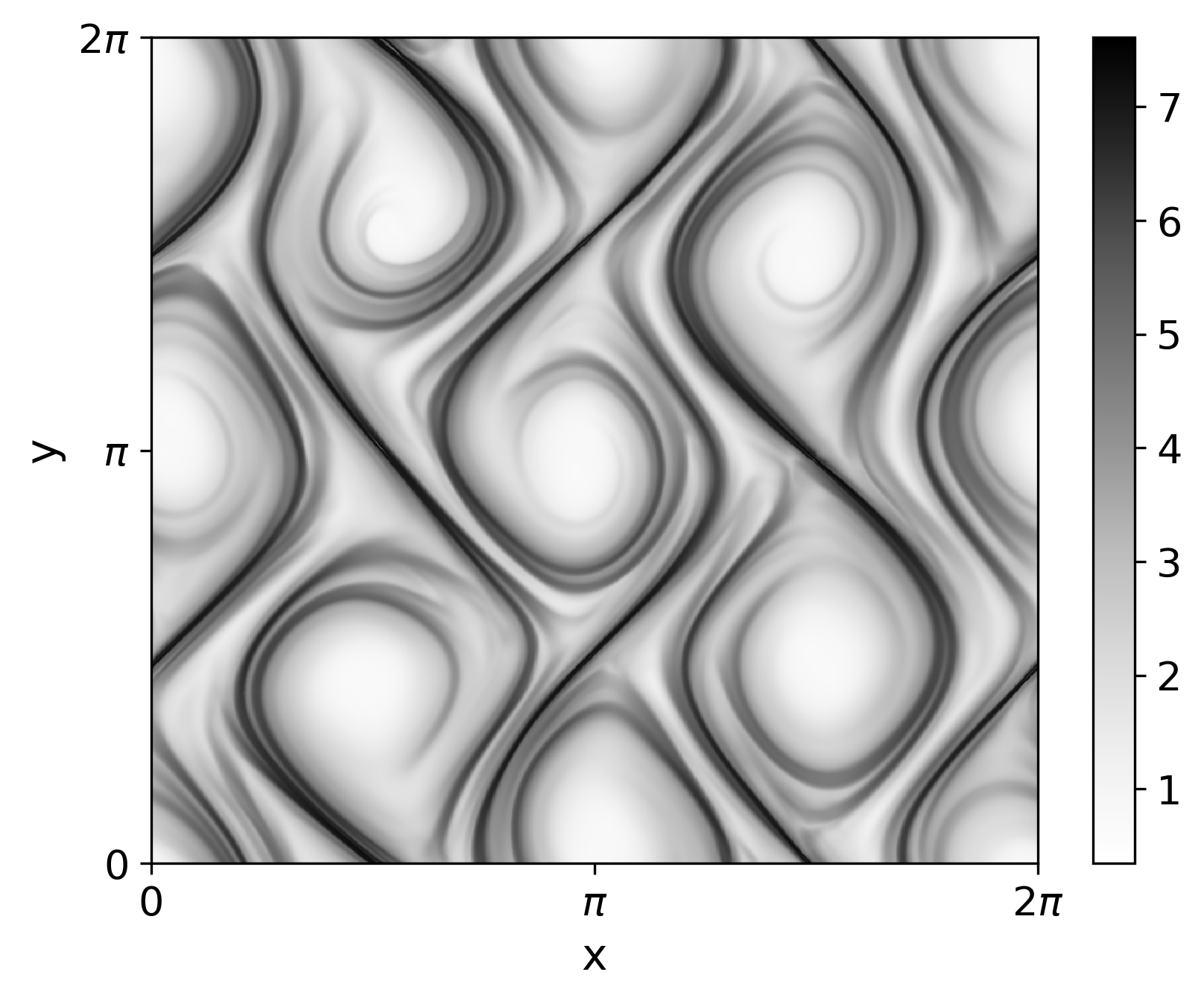}
\caption{FENE-P model:
Representative snapshots of $\ln(\operatorname{tr}\mathsfbi{C})$
in the steady state for ${\it Sc}=10^3$ (left) and  ${\it Sc}=\infty$ (right).
}
\label{fig:conformation-fene-p}
\end{figure}

In this Appendix, we report
simulations of the FENE-P model with and without artificial diffusion.
The maximum polymer square extension is $\ell^2_{\rm max}=3\times 10^3$, while
the other parameters of the FENE-P model are the same as those used for
the Oldroyd-B model in \S~\ref{sect:results}. Equation~\eqref{eq:C-FENE-P}
is solved by combining the Cholesky decomposition and the Kurganov-Tadmor
hyperbolic solver as described in \S~\ref{sect:model}. The effect of
artificial diffusivity is once again studied by adding the term $\kappa\Delta\mathsfbi{C}$
with $\kappa=5\times 10^{-5}$ to \eqref{eq:C-FENE-P}, so that $\mathit{Sc}=10^3$.
Figures~\ref{fig:spectra-fene-p} and~\ref{fig:conformation-fene-p}
are the counterparts for the FENE-P model
of figures~\ref{fig:kinetic} (right panel), \ref{fig:stress} 
and \ref{fig:spectra} (left panel). The results show that the 
conclusions on the effects of artificial diffusivity obtained for the
Oldroyd-B model hold analogously for the FENE-P model.

\section*{Supplementary material}

# Effect of polymer-stress diffusion in the numerical simulation of elastic turbulence

# SUPPLEMENTARY MATERIAL

## Anupam Gupta[1] and Dario Vincenzi[2]

[1]Paulson School of Engineering and Applied Sciences, Harvard University, Cambridge, Massachusetts 02138, USA

[2]Université Côte d'Azur, CNRS, LJAD, 06100 Nice, France

This supplementary material consists of additional simulations of the Oldroyd-B model on the periodic square $[0, 2\pi]^2$ with a cellular forcing (see (2.1) and (2.2) in the main text). The integration method is described in § 2 of the manuscript.

Figures 1 and 2 are the analogues of figures 2 (right panel), 6 and 8 (left panel) in the main text for $\tau = 40$, $\mu = 0.02$ and same $K$, $f_0$, $\nu$ as in § 2. In particular, the Deborah number is now $De = 8$ and the coupling coefficient $\mu$ is doubled.

Figures 3 and 4 are the analogues of figures 2 (right panel), 6 and 8 (left panel) in the main text for $K = 4$, $\tau = 10$, $f_0 = 0.16$ and same $\nu$ and $\mu$ as in § 2. The Deborah number is $De = 8$ and the spatial frequency of the forcing is doubled compared to the main text.

These additional simulations support the validity of the conclusions reported in the manuscript.

Finally, figure 5 shows the quantity $d^2(\mathbf{I}, \mathbf{C}) = \mathrm{tr}\log^2(\mathbf{C})$, which represents the geodesic distance between the conformation tensor and the identity tensor in the space of positive-definite tensors [see Hameduddin *et al.*, *J. Fluid Mech.* **842**, 395 (2018)] for the same parameters as in § 3 of the main text.

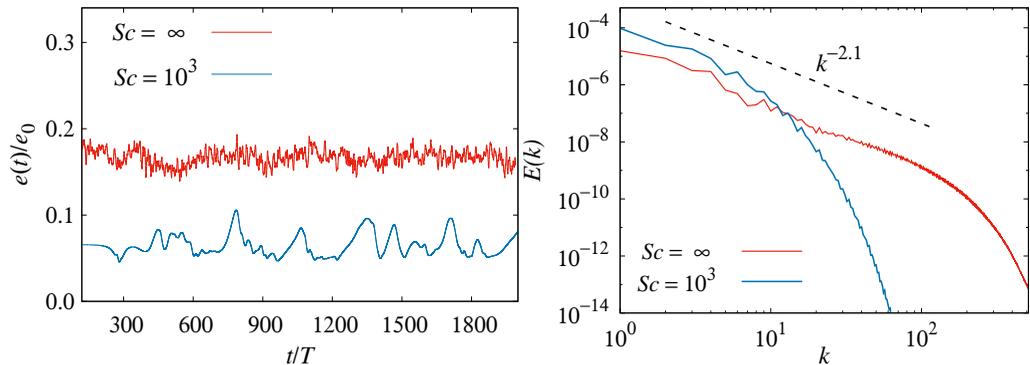

Fig. 1. Time series of the rescaled kinetic energy in the steady state (left) and kinetic-energy spectra (right) for the Oldroyd-B model with $\nu = 0.05$, $f_0 = 0.02$, $K = 2$, $De = 8$, $\mu = 0.02$.

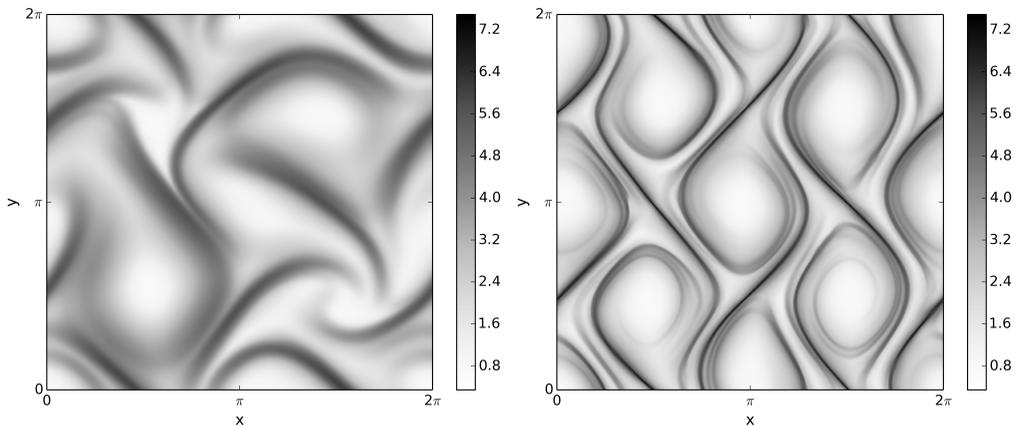

Fig. 2. Representative snapshots of $\ln(\mathrm{tr}\,\boldsymbol{C})$ in the steady state for the Oldroyd-B model with $\nu = 0.05$, $f_0 = 0.02$, $K = 2$, $De = 8$, $\mu = 0.02$ and $Sc = 10^3$ (left) and $Sc = \infty$ (right).

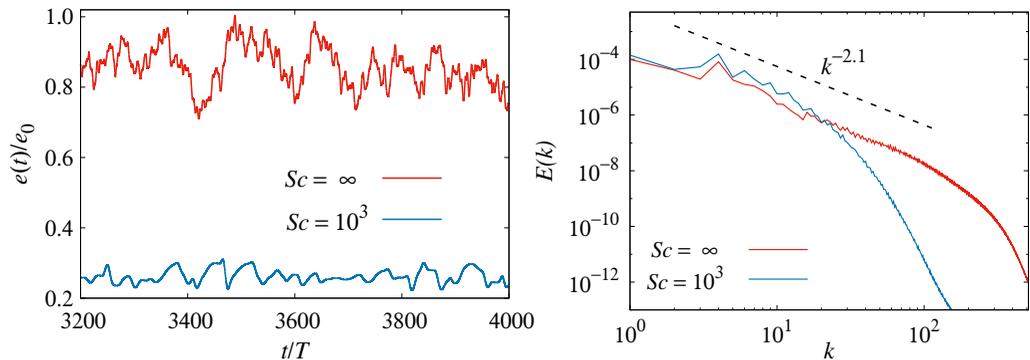

Fig. 3. Time series of the rescaled kinetic energy in the steady state (left) and kinetic-energy spectra (right) for the Oldroyd-B model with $\nu = 0.05$, $f_0 = 0.16$, $K = 4$, $De = 8$, $\mu = 0.01$.

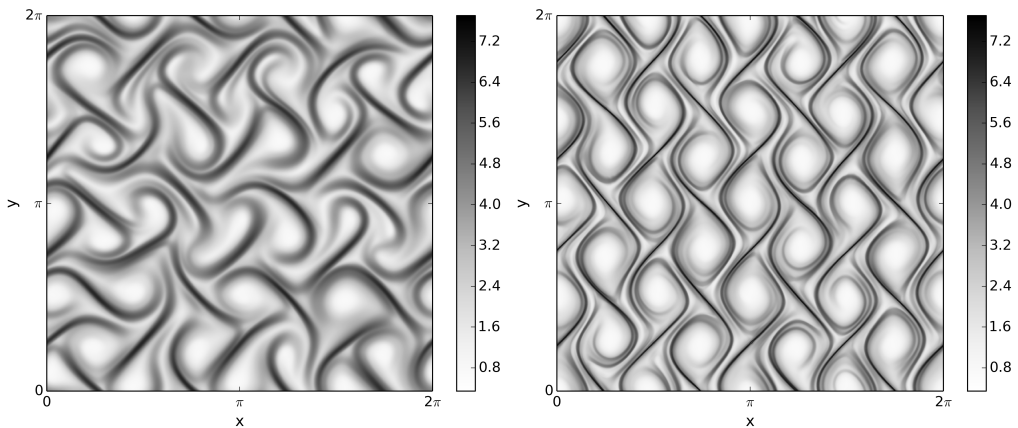

Fig. 4. Representative snapshots of $\ln(\mathrm{tr}\,\boldsymbol{C})$ in the steady state for the Oldroyd-B model with $\nu = 0.05$, $f_0 = 0.16$, $K = 4$, $De = 8$, $\mu = 0.01$ and $Sc = 10^3$ (left) and $Sc = \infty$ (right).

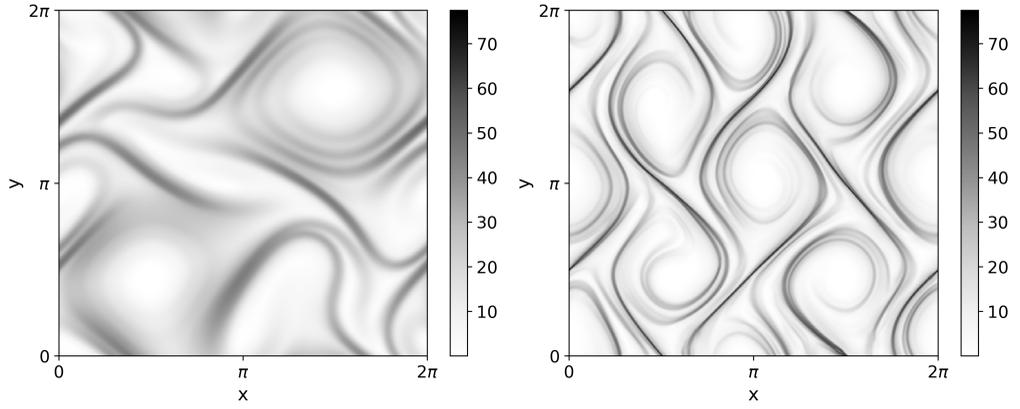

Fig. 5. Snapshots of $d^2(\mathbf{I}, \mathbf{C})$ for $Sc = 10^3$ (left) and $Sc = \infty$ (right) and the same parameters and time instants as in figure 6 of the main text.



\end{document}